\documentclass[aip,reprint,onecolumn]{revtex4-1}

\usepackage{graphicx}
\usepackage{dcolumn}
\usepackage{bm}
\usepackage{color}

\begin{document}

\title{\centering{Negative experimental evidence for magneto-orbital dichroism \\ Supplemental information}}
 
 \vspace{1cm}
\author{Renaud Mathevet}
\email{renaud.mathevet@lncmi.cnrs.fr}
\affiliation{CNRS-INSA-UJF-UPS, Laboratoire National des Champs Magn\'etiques Intenses,\\ F–31400 Toulouse, France}
\affiliation{Universit\'e de Toulouse, LNCMI-T, F-31062 Toulouse, France}
\author{Bruno Viaris de Lesegno}
\author{Laurence Pruvost}
\affiliation{Laboratoire Aim\'e Cotton, CNRS/Univ Paris-Sud/ENS-Cachan, B\^at 505, \\ Campus d'Orsay, F-91405 Orsay Cedex, France}
\author{Geert L. J. A. Rikken}
\affiliation{CNRS-INSA-UJF-UPS, Laboratoire National des Champs Magn\'etiques Intenses,\\ F–31400 Toulouse, France}
\affiliation{Universit\'e de Toulouse, LNCMI-T, F-31062 Toulouse, France}

\vspace{1cm}

\begin{abstract}
A light beam can carry both spin angular momentum (SAM) and orbital angular momentum (OAM). SAM is commonly evidenced by circular dichroism (CD) experiments {\em i. e.} differential absorption of left and right-handed circularly polarized light. Recent experiments, supported by theoretical work, indicate that the corresponding effect with OAM instead of SAM is not observed in chiral matter.\\
Isotropic materials can show CD when subjected to a magnetic field (MCD). In Ref.~\onlinecite{Mathevet2012} we report a set of experiments, under well defined conditions, searching for magnetic orbital dichroism (MOD), differential absorption of light as a function of the sign of its OAM. We experimentally demonstrate that this effect, if any, is smaller than a few $10^{-4}$ of MCD for the Nd:YAG $^4I_{9/2}\rightarrow^4F_{5/2}$ transition. This transition is essentially of electric dipole nature. We give an intuitive argument suggesting that the lowest order of light matter interaction leading to MOD is the electric quadrupole term. \\
We give here more experimental details and extra measurements. 
\end{abstract}


\maketitle

\section{Details on the experimental setup}
\label{Sec:ExpSetup}

\begin{figure}[htbp]
\includegraphics[width=8.5cm]{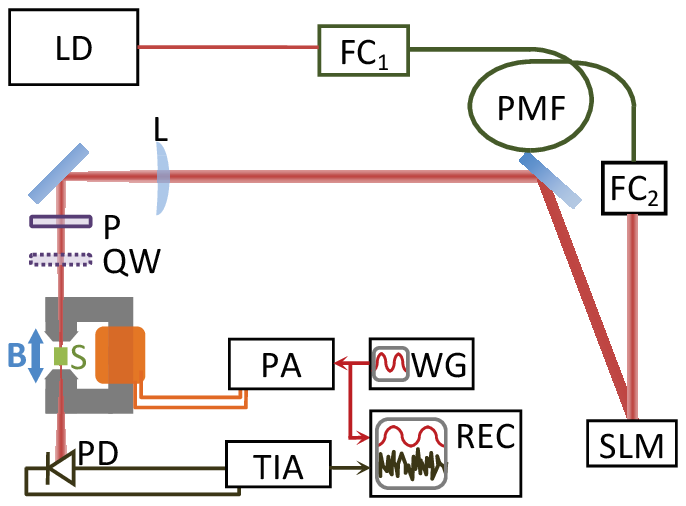}\hfill
\includegraphics[width=8.5cm]{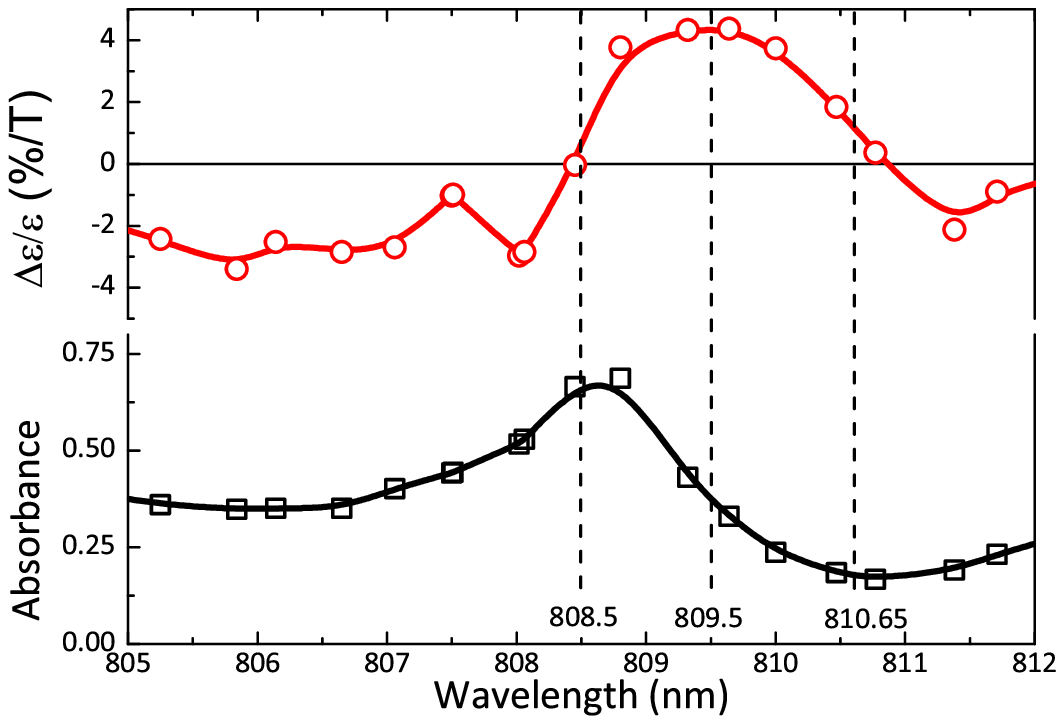}
\caption{Left: experimental Setup. LD: laser diode, FC: fiber couplers, PMF: polarization maintaining monomode fiber, SLM: spatial light modulator, L: lens, P: polarizer, QW: quater-wave plate, S: sample, B: AC longitudinal $B-$field, PD: photodiode, WG: waveform generator, TIA: transimpedance amplifier, PA: power amplifier, REC: recorder. Right: Absorbance (black) and MCD (red) spectra of Nd:YAG around $809~\mathrm{nm}$.}
\label{fig:setup}
\end{figure}

The experimental setup is depicted on Fig.~\ref{fig:setup}-left. Light from a $200~\mathrm{mW}$, $808~\mathrm{nm}$ laser diode (Radiospares DL-808-0.2) is coupled into a $10~\mathrm{m}$-long polarization maintaining monomode fiber for spatial mode filtering. This diode is longitudinally and spatially multimode. It results in a very poor coupling ($5\%$) but we found lower overall noise than with a monomode laser diode (Thorlabs L808P010 or L808P030).

Light emerging from a $f=36~\mathrm{mm}$ outcoupler (Thorlabs F810FC-780) is typically $6~\mathrm{mm}$ in diameter and directed onto a Spatial Light Modulator (Hamamatsu LCOS X10468-02). A blazed grating is superimposed on the helical phase map to be imprinted on the wavefront. The desired Laguerre-Gaussian beam is then diffracted at a 2~mrad angle from the reflected beam which is subsequently easily blocked.

The Laguerre-Gaussian beam is then slightly focussed by a $f=750~\mathrm{mm}$ lens. The beam waist is $w_0=78~\mathrm{\mu m}$ and is located $\sim40~\mathrm{mm}$ behind the sample. It corresponds to a beam divergence $\theta_0=3.3~\mathrm{mrad}$. Propagation of a converging beam mixes the OAM with SAM. According to~\cite{Nieminen2008} the coupling strength is $\theta_0^2/4\sim 3\times 10^{-6}$, negligible compared to the $0.1$ value used in~\cite{Loffler2011}.

The sample is a $3~\mathrm{mm}$ in diameter, $2~\mathrm{mm}$-long Nd:YAG rod at a concentration of $\sim1~$at.$\%$ (MolTech GmbH).

The last optical element before the sample light is a polariser (Thorlabs LPVIS050-MP: extinction ratio $>10^7$). Light is then collected on a $\sim(4~\mathrm{mm})^2$ Si-PIN photodiode (Thorlabs FDS100 with front window removed) whose photocurrent is amplified by a low noise transadmittance amplifier (Stanford Research SR570, low noise mode, $30-300~\mathrm{Hz}$ bandpass filter). All these elements are placed at good distance, typically $\sim40~\mathrm{cm}$, from the electromagnet.

The sample is located in the $\sim3~\mathrm{mm}$ gap of an electromagnet built from a transformer. It produces a typical $B=330~\mathrm{mT_{RMS}}$ for a current $I=2.75~\mathrm{A_{RMS}}$ at $f_{mod.}=85.75~\mathrm{Hz}$ supplied by a bipolar power amplifier (Kepco BOP36-12M). Impedance at the working frequency is lowered by use of a series $100~\mu\mathrm{F}$ capacitor shorted by a $8.2~\mathrm{k}\Omega$ resistor to avoid over charge by offset DC currents.

The modulation and photodiode signals are finally fed into a recorder (Hioki 8860 with a 8957 HiRes unit) for subsequent computer manipulation.

For the sake of quantitative comparison, a conventional MCD experiment can be performed placing an achromatic quarter-wave plate (Thorlabs AQWP05M-980) just after the polarizer at $\pm 45^\circ$ from the polarization direction and using a $LG_0$ beam.

The incident power on the sample is on the order of $2.6~\mathrm{mW}$. Measured transmission at $809.5~\mathrm{nm}$ is $T=43~\%$ and MCD coefficient~$4.2\%/\mathrm{T}$. The differential transmitted power amplitude is thus $7.7~\mathrm{\mu W}$ which amounts to $0.3\%$ of the incident power. With Si-PIN Photodiode responsivity of about $0.5~\mathrm{A/W}$ and a transadmittance of $20~\mathrm{\mu A/V}$ we record a MCD signal whose amplitude is $190~\mathrm{mV}_{RMS}$ typically.

\section{Prior alignement}
\label{Sec:alignement}

We found that special care must be taken to alignement. If the sample is slightly tilted with respect to the laser beam, Fresnel coefficients at the entrance side are different in amplitude and phase. This might result in a small circularly polarized component propagating in the sample which, in turn, is subjected to a comparatively strong MCD effect.

In the same way, we made a stiff, non-magnetic holder (glass fiber composite) and aluminum posts. The sample is tightened with a Nylon screw. The pressure exerted might induce some birefringence which converts the incoming linear polarization into an elliptic one subjected to strong MCD effect. The sample is set as loosely as possible. A local residual birefringence of the sample cannot be excluded too. As a consequence, the polarizer direction must be tuned to minimize the signal recorded with a $LG_0$ beam at the modulation frequency. Under the worse positionning/alignement conditions, we found this unwanted effect to give a signal $30$ times higher than the noise level of the experiment.

Any experiment presents some drifts in particular here associated with heating from the electromagnet. This is the reason why $B-$field was deliberately kept to a third of the maximum value we can obtain with our supply. Unfortunately, it also makes systematic studies with best resolution, {\em i. e.} long acquisition times, very cumbersome.

\section{$LG$ beams obtained with an SLM}
\label{Sec:Photos}
\begin{figure}[htbp]
\centering
\includegraphics[width=17cm]{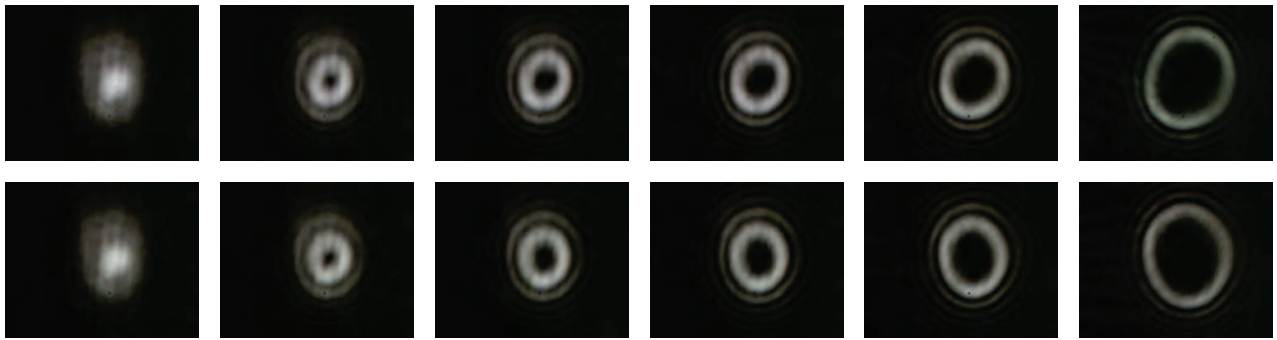}
\caption{Pictures of the different $LG_\ell$ beams used. Upper row: $\ell=0,+1,+2,+3,+5,+10$. Lower row: $\ell=0,-1,-2,-3,-5,-10$.}
\label{fig:photos}
\end{figure}
We present in Fig.~\ref{fig:photos} some pictures recorded on a simple webcam with lens removed. Its response has been deliberately made non linear (gamma correction and contrast settings to their maximum value) to enhance imperfections such as fringes corresponding to non zero values of the radial index $p$ of the $LG_\ell^p$ expansion basis. However, these imperfections do not affect our experiment. Indeed, as long as the phase helix imprinted on the wavefront has a regular pitch and an $\ell\;2\pi$ maximum phase shift, the expansion on the Laguerre-Gaussian modes basis is limited to that single value of $\ell$. The emerging beam has thus a well defined OAM.

\section{Spectra for different wavelengths}
\label{Sec:DiffLambdas}

\begin{figure}[htbp]
\includegraphics[width=8.5cm]{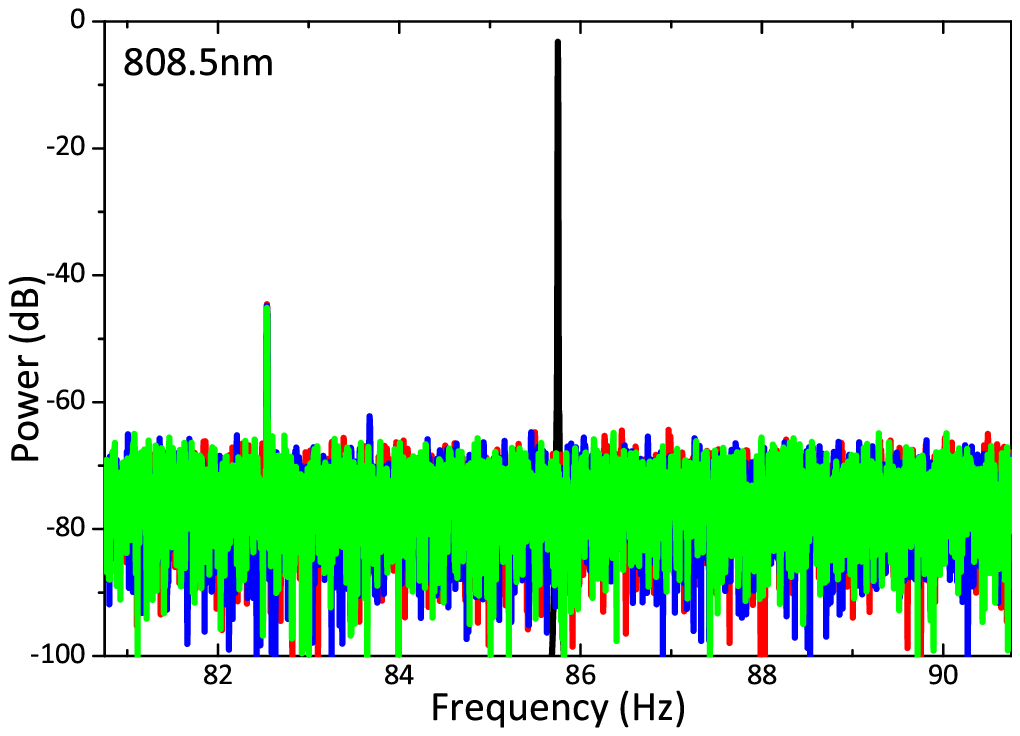}\hfill
\includegraphics[width=8.5cm]{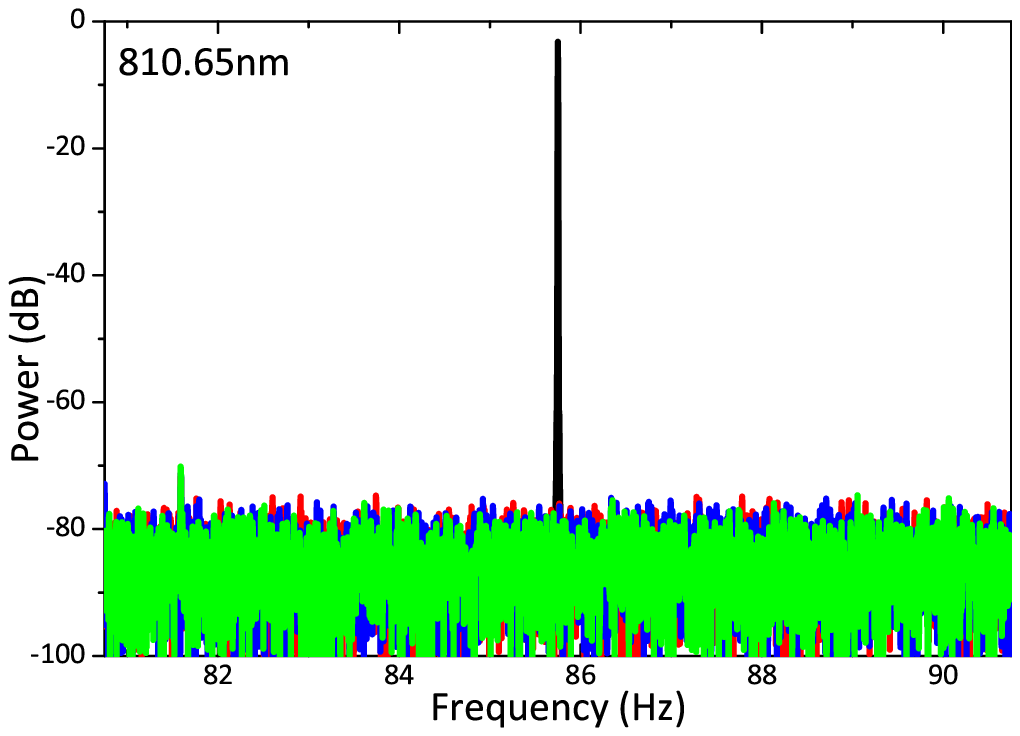}
\caption{Spectra recorded for different wavelengths for which MCD signal is 0. The black curve is the reference, maximum MCD signal at $809.5~\mathrm{nm}$. Red/blue/green corresponds respectively to $\ell=-1/0/+1$. Left: $808.5~\mathrm{nm}$ corresponds to a maximum of absorption. $810.65~\mathrm{nm}$ corresponds to a minimum of absorption.}
\label{fig:DiffLambda}
\end{figure}

The lineshape of the MCD is not clearly absorption-like or dispersion-like (see Fig.~\ref{fig:setup}-right). Anyway, MOD involves different molecular symmetries and could have, for the same transition, a different lineshape. By chance, it could happen that MOD is (almost) null when MCD is maximum. We thus checked on both sides of the maximum if any signal could be recorded. As seen on Fig.~\ref{fig:DiffLambda} no such a signal was found. It can be noticed that for the $810.65~\mathrm{nm}$ spectrum the laser noise floor is lower (see Sec.~\ref{Sec:LaserNoise}).

\section{Spectra from $LG_{-10}$ to $LG_{+10}$}
\label{Sec:DiffLGs}

\begin{figure}[htbp]
\centering\includegraphics[width=8.5cm]{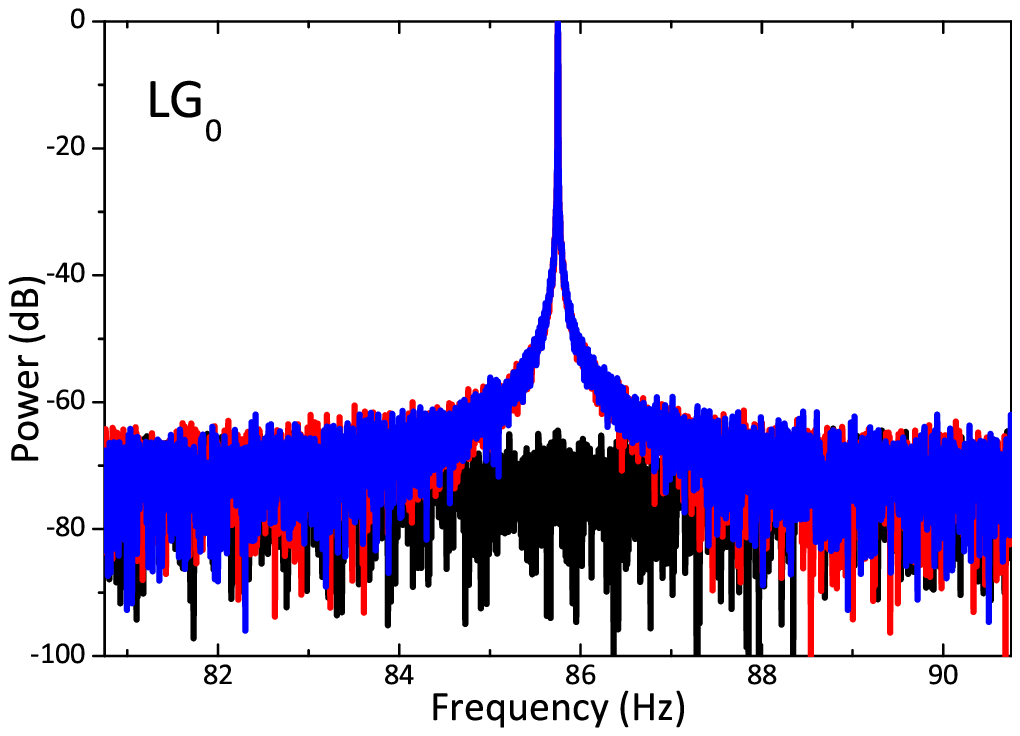}\hspace{6.5cm}
\includegraphics[width=8.5cm]{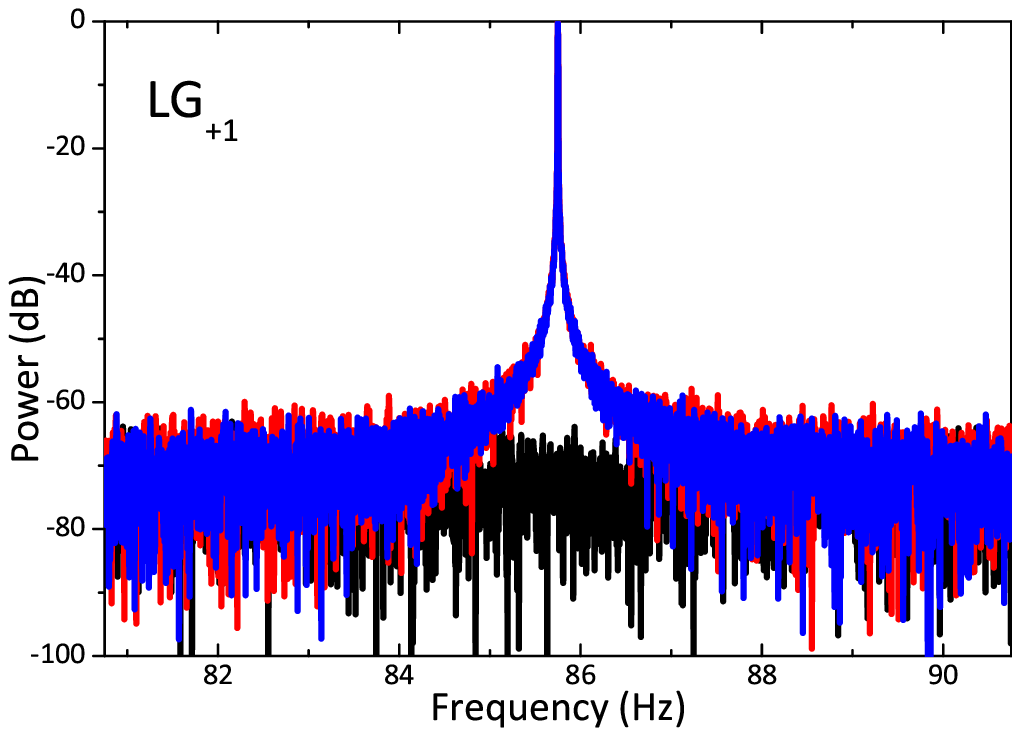}\hfill
\includegraphics[width=8.5cm]{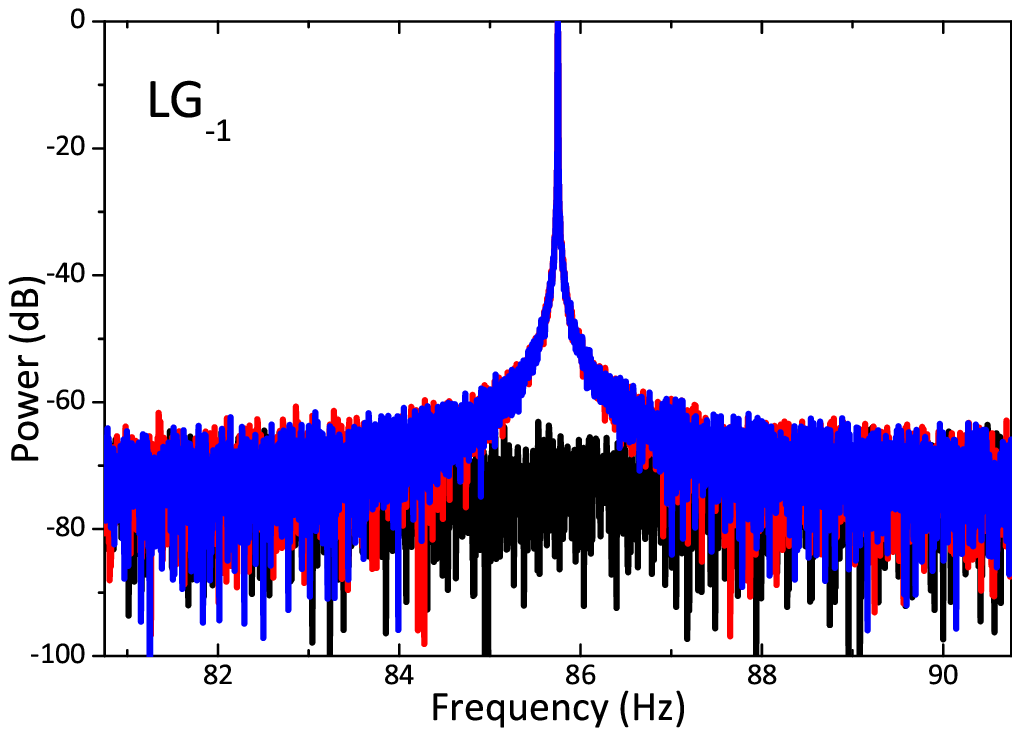}
\includegraphics[width=8.5cm]{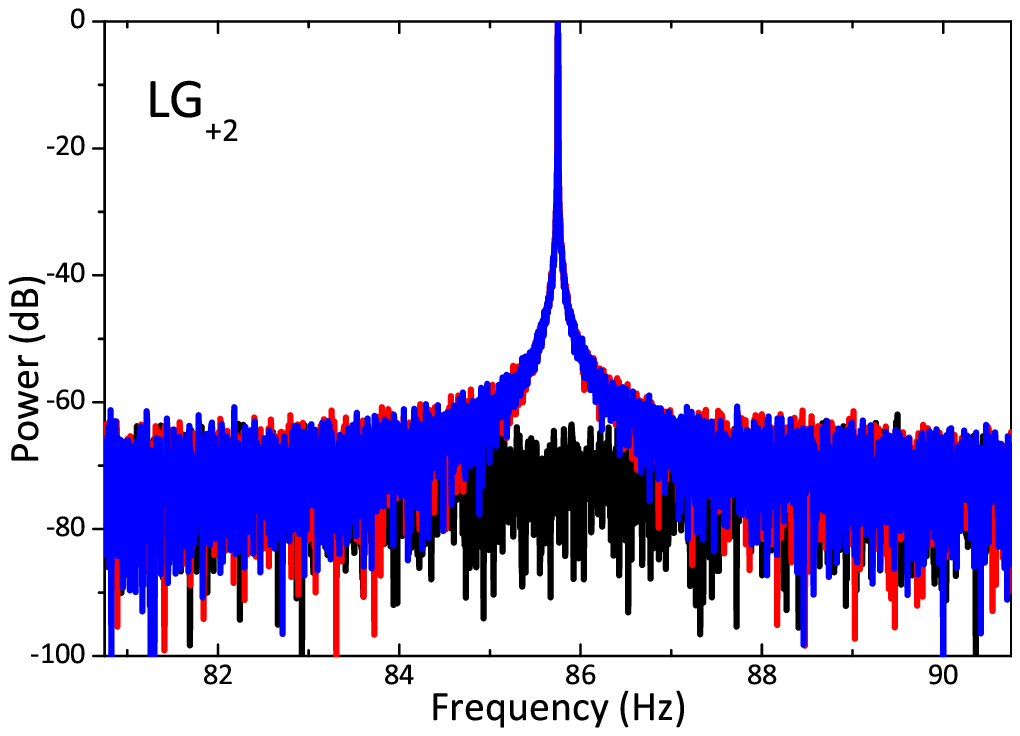}\hfill
\includegraphics[width=8.5cm]{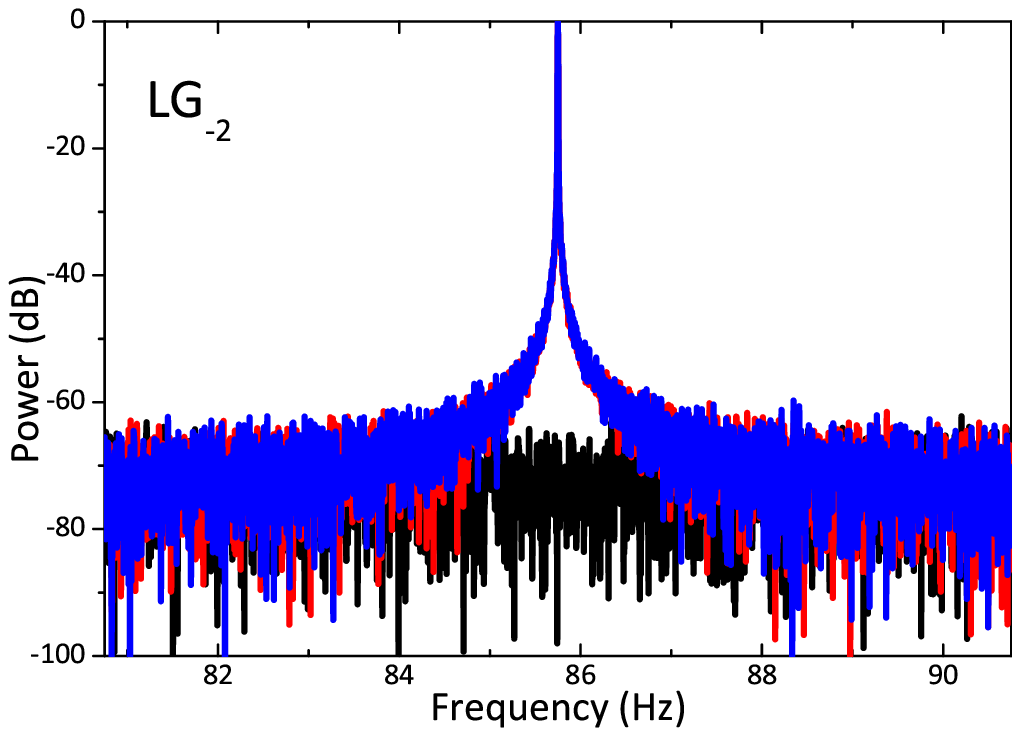}
\caption{Spectra recorded for different $LG_\ell$'s beams corresponding to $\ell\hbar$ units of OAM. Black: linearly polarized light (no SAM), Red/Blue circularly polarized light of opposite helicities ($\pm\hbar$ units of SAM).}
\label{fig:AllLGs}
\end{figure}

We present in Fig.~\ref{fig:AllLGs} the spectra obtained at $809.5~\mathrm{nm}$ where MCD is maximum for different $LG_\ell$ beams for $\ell$ values ranging from $-2$ to $+2$ for three different angular momentum configurations. With no quarter-wave plate after the polarizer, light is linearly polarized and (SAM, OAM) corresponds to $(0\hbar$, $\ell\hbar)$. This pure OAM configuration is depicted in black. A quarter-wave plate is then set after the polarizer at $\pm45^\circ$ from its polarization direction. Light is circularly polarized and (SAM, OAM) corresponds to ($\pm 1\hbar$, $\ell\hbar$). This corresponds to a mixed configuration that fixes the scale of a reference MCD signal (Red/Blue). In all cases, MCD signals are equal and no MOD signal is found above the noise floor. Due to acquisition parameters and shorter integration time (see Sec.~\ref{Sec:alignement}), spectral resolution and sensitivity are lower here than for the data presented in the main article.

\begin{figure}[htbp]
\includegraphics[width=8.5cm]{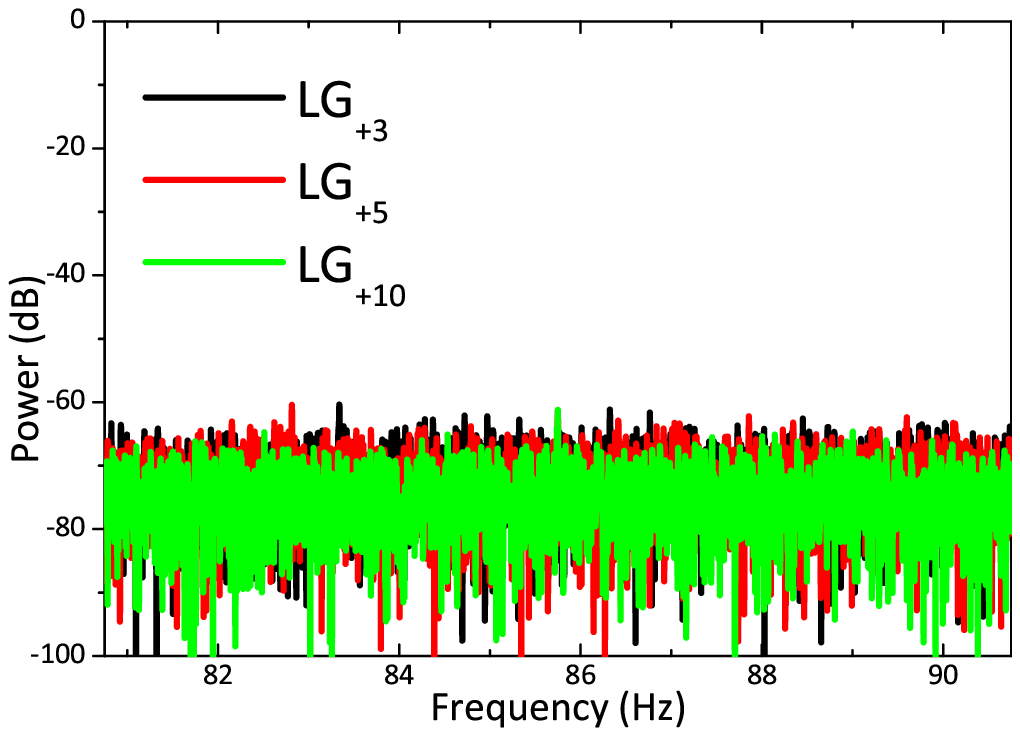}\hfill
\includegraphics[width=8.5cm]{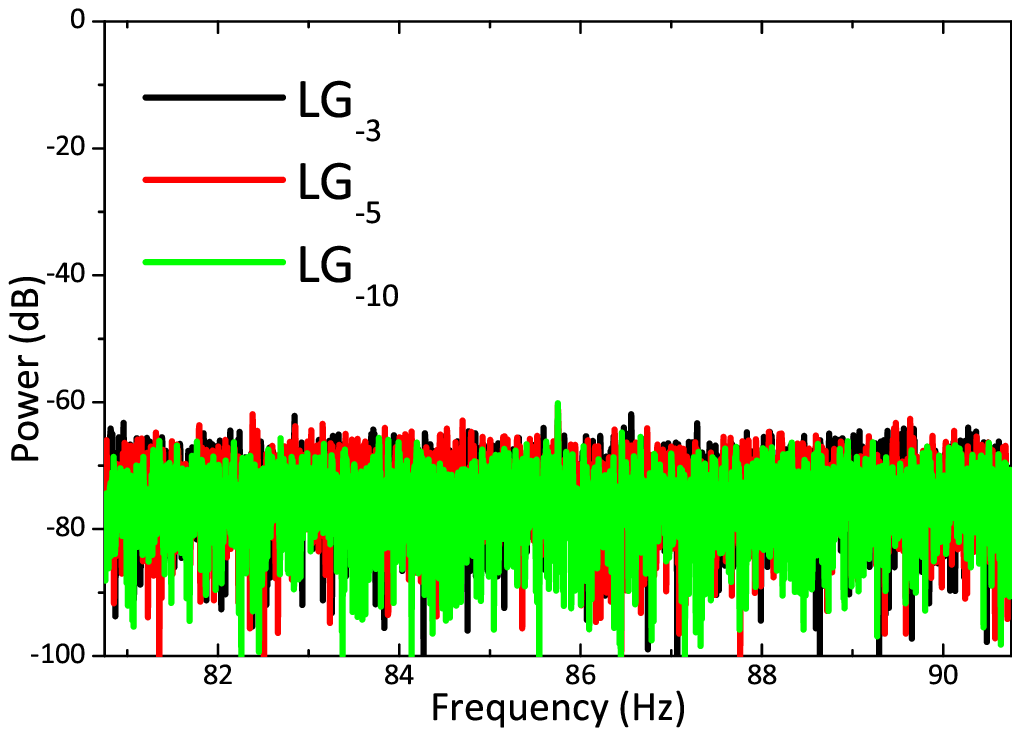}
\caption{Spectra recorded for different $LG_\ell$'s beams corresponding to an OAM $\ell\hbar$. Color code on the graph.}
\label{fig:LGSup}
\end{figure}

In Fig.~\ref{fig:LGSup} we show spectra for higher values of $|\ell|$ but only in the pure OAM configuration: color code now distinguishes the different values of $\ell$. Here again no evidence of MOD is found. As can be noticed, the noise density is slightly lower for the experiments with $LG_{\pm 10}$ beams. This shouldn't be misinterpreted. As $|\ell|$ increases, the mode spatial extension grows (see Fig.~\ref{fig:photos}). For $|\ell|\geq10$ the aperture drilled in the electromagnet blocks part of the beam and the overall intensity is reduced. This is why we restricted ourselves to $|\ell|\leq10$. Besides, one can hardly imagine an elementary process involving more than a few $\hbar$ of angular momentum.

\section{Laser noise}
\label{Sec:LaserNoise}

\begin{figure}[htbp]
\centering\includegraphics[width=8.5cm]{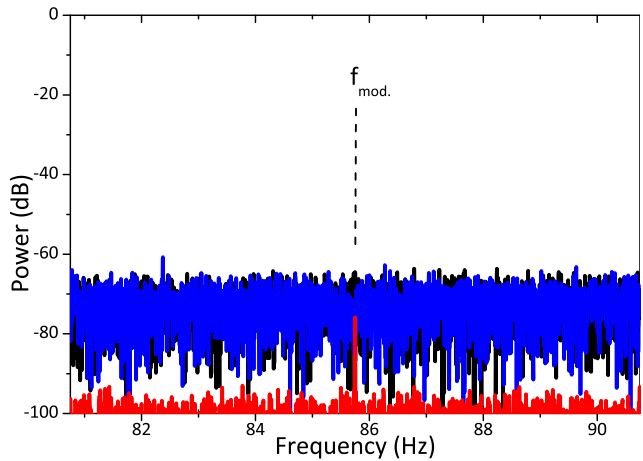}
\caption{Noise spectrum in the $f_{mod.}=85.75~\mathrm{Hz}$ region. Black/Blue: Laser On, $B-$field on/off. Red: Laser off, $B-$field on.}
\label{fig:noise}
\end{figure}

The laser diode is driven by an homemade low-noise power supply originally designed for laser diode spectroscopy. The noise level of our experiment is dominated by laser intensity noise in the $85.75~\mathrm{Hz}$ region. We check it comparing spectra acquired with an $LG_0$ beam and $B-$field on and off (Fig.~\ref{fig:noise}). The analysis bandwidth is made of 4 bins of $0.4~\mathrm{mHz}$ that is $BW=1.6~\mathrm{mHz}$. At $809.5~\mathrm{nm}$, the power ratio of the energy in analysis band to the MCD signal is $\eta_0=1.2\times 10^{-7}$ (see main article). The MCD signal itself is $x=0.3\%$ of the $2.6~\mathrm{mW}$ incident power.

It can be noticed that this noise not only comes from intensity noise of the laser source. When the input polarization of the beam is not perfectly aligned with the polarization axis of the fiber, the output polarization is slightly elliptical. As a consequence polarization noise is converted into intensity noise after subsequent polarizers. This might be the reason why the noise floor at different wavelengths is different (Fig.~\ref{fig:DiffLambda}).

On Fig.~\ref{fig:noise} is also plotted the electronic noise spectrum (in red) obtained when laser light is blocked before coupling into the fiber. One clearly sees a well defined peak at the modulation frequency whose amplitude is only slightly lower than the laser noise level. However, its phase is in quadrature with the $B-$field modulation. It thus corresponds to electronic pick-up ($\propto \partial B/\partial t$) that can thus be distinguished from the actual in phase signal ($\propto B$) by phase sensitive detection (see main article).

\section{Allan variance analysis}
\label{Sec:Allan}

\begin{figure}[htbp]
\includegraphics[width=8.5cm]{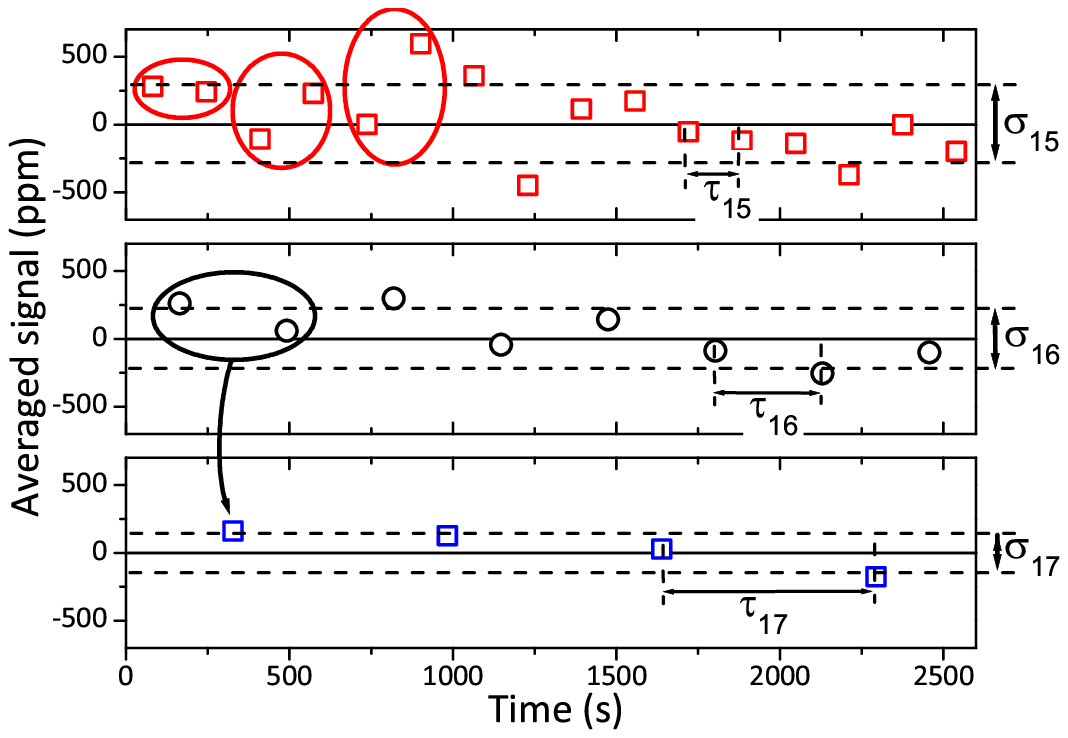}\hfill
\includegraphics[width=8.5cm]{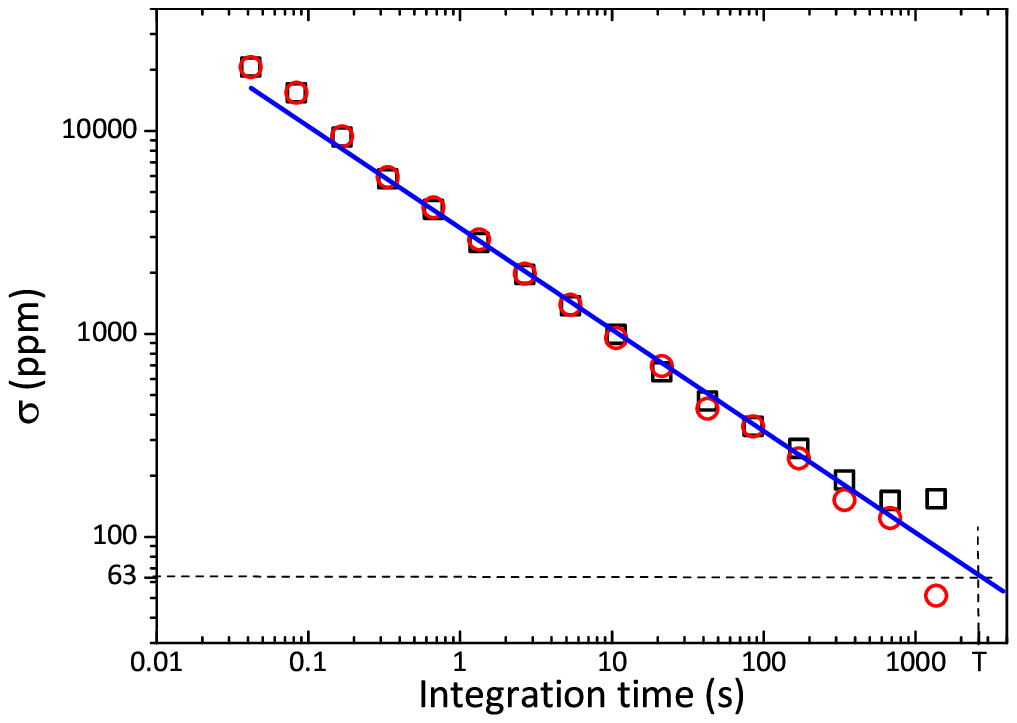}
\caption{Principle of Allan variance analysis. Left: three iterative steps. At each step two samples of the previous step are averaged to simulate a twice longer integration time. At each step a variance on the set of samples is evaluated. Right: plot of the variance as a function of the integration time. Red circles/ black squares corresponds to the in-phase and in-quadrature signals with the $LG_1$ beam. Blue line is a $\tau^{-1/2}$ fit. Deviation from this line for the two last points is irrelevant (see text).}
\label{fig:Allan}
\end{figure}

The data for the $LG_0$ and $LG_1$ beams presented in the main article correspond to single realization of a random processes. Direct comparaison of the two values has low significance and is to be interpreted relative to the dispersion of individual results.

To evaluate such a dispersion from a single run we perform an Allan variance analysis. In a given temporal series, each individual sample corresponds to a $\tau$ integration time. We can calculate the variance $\sigma_{1\tau}$ over the whole set of $N$ samples. Then we compute the mean of each pair of two successive samples. We get a set of $N/2$ samples simulating a $2\tau$ integration time on which the variance $\sigma_{2\tau}$ is evaluated. The procedure is repeated recursively. In Fig.~\ref{fig:Allan}-left, we show three of such iterations. As can be seen, at step $17$ there is only $4$ samples left and the two next steps will have only $2$ and $1$ sample. As a consequence, the associated variance is not really reliable as can be seen on Fig.~\ref{fig:Allan}-right: the two last point deviate from the classical inverse square root dependence of the variance with respect to the simulated integration time (blue line). The measurement is the mean value over the full set of samples. It corresponds to a $T$ integration time at which the variance is extrapolated to $\sigma_T=63~\mathrm{ppm}$. On a separate longer acquisition time series we checked that inverse square root law was still valid for the actual experiment acquisition time. Extrapolation is thus legitimate.

\section*{\label{Bibliography} Bibliography}

\end{document}